\documentclass{article}
\pdfoutput=1
\usepackage[utf8]{inputenc}
\usepackage{amsmath}
\usepackage{amssymb}
\usepackage[margin=1in]{geometry}
\usepackage{graphicx}
\usepackage{algorithm}
\usepackage{algorithmic}
\usepackage{hyperref}
\usepackage{verbatim}
\usepackage{float}
\usepackage{multirow}
\usepackage{color}
\usepackage{subcaption}
\usepackage{csquotes}
\usepackage[backend=biber,style=nature]{biblatex}
\hypersetup{
	unicode={true},
	pdfstartview={FitH},
	bookmarksnumbered={true},
	bookmarksopen={true},
	pdfborder={0 0 0},
	colorlinks=false
}
\usepackage{abstract}
\usepackage{indentfirst}
\usepackage[affil-it]{authblk}
\usepackage[english]{babel}
\usepackage{blindtext}
\usepackage{threeparttable}

\title{Resampling community detection to maximize propagation in complex networks}

\date{}
\author{Xintong Zhai}
\author{Zhonghao Xu\thanks{Corresponding author. Email: zhxu@sfs.ecnu.edu.cn.}}
\affil{School of Statistics, East China Normal University}

\begin{document}
\maketitle
\vspace*{-1.2cm}
\renewcommand{\abstractnamefont}{\normalfont\Large\bfseries}
\renewcommand{\abstracttextfont}{\normalfont}

\begin{abstract}
    Identifying important nodes in complex networks is essential in theoretical and applied fields. A small number of such nodes have deterministic power to decide information spreading, so it is of importance to find a set of nodes that maximize the propagation in networks. Based on baseline ranking methods, various improved methods were proposed, but there does not exist one enhanced method that covers all the base methods. In this paper, we propose a penalized method called RCD-Map, which is short for resampling community detection to maximize propagation, on five baseline ranking methods(Degree centrality, Closeness centrality, Betweennees centrality, K-shell and PageRank) with nodes' local community information. We perturbed the original graph by resampling to decrease the biases and randomness brought by community detection methods – both overlapping and non-overlapping methods. To assess the performance of our identifying method, SIR(susceptible-infected-recovered) model is applied to simulate the information propagation process. The result shows that methods with penalties perform better with a vaster propagation range in general.

\end{abstract}

\section{Introduction}

Complex network is one of the most important tools in researching complex systems. It has been part of human beings' production and daily life. Ecological systems, power grid, social relationship networks and financial markets can all be considered as complex systems. Trace back to the 1960s, the study of complex networks was initialed by two mathematicians Erd{\H{o}}s and R{\'e}nyi, who founded systematic research about random graph \cite{erdHos1960evolution}. In 1998, small-world property was uncovered by Watts and Strogatz \cite{watts1998collective}, and then in 1999, scale-free property was discovered by Barab{\'a}si and Albert \cite{barabasi1999emergence}. These three researches are the basis of complex network study, and the latter two also bridged the gap between theory and reality. Applying complex networks to model real systems can reveal the properties of these complex systems efficiently. 

Nodes are major components of complex networks, mining important nodes is one of the fundamental aspects of complex network research. Important nodes are referred to as certain nodes with more decisive influence on the structure and function of the network compared with other nodes. Despite the limited amount, these special nodes have great propagation ability. Identifying important nodes not only has theoretical meanings but also owns reality value. For example, blogs published by key opinion leaders(KOL) will soon sweep across the Internet; a handful of broken high tension cables in Ohio could result in a large-scale outage in North America. Therefore, sorting and mining important nodes has important meanings in a broad range of applications \cite{lu2016vital}. 

There are two directions for identifying important nodes: one direction is to rank individual nodes, the other is to find out a set of the most influential nodes. The mainstream methods of the former problem are based on three types of centrality measures with different priorities: neighborhood-based centrality, path-based centrality, and eigenvector-based centrality\cite{lu2016vital,xiaolong2014review}. Neighborhood-based centrality includes the famous degree centrality(DC) \cite{bonacich1972factoring}, which is fast and direct to compute but it neglects location information. K-shell decomposition \cite{kitsak2010identification} locates the nodes in the network with coarse-grained scores. Path-based centrality includes closeness centrality(CC) \cite{freeman1978centrality}, betweenness centrality(BC) \cite{bavelas1948mathematical},and Katz centrality \cite{katz1953new}. These methods traverse all paths in the network so they suffer from high computation complexity. Eigenvector-based methods not only considers the number of neighbors, but also calculates the influence of their weights. While Eigenvector centrality \cite{bonacich1972factoring} is applied to undirected networks, other iterative algorithms like PageRank \cite{brin1998anatomy} and LeaderRank \cite{lu2011leaders}, the improved version with a background node, are used in directed networks. In the field of identifying a set of vital nodes, there also exist heuristic and greedy algorithms \cite{lu2016vital} such as VoteRank algorithm that can select vital nodes one by one via voting, and at each timestep, the voting ability of every neighbor node of the selected node will be penalized \cite{zhang2016identifying}, however, the penalty of VoteRank is set to be a constant, the inverse of network's average degree, which is unreasonable. Overall, most of the promoted algorithms can only be seen as the improved version of one centrality. There lacks an improvement ranking method suitable to all centrality measures.

Networks' structural properties, like community structures helps to cope with certain problems in complex networks. Since Wasserman discovered the community structure of complex networks in 1994 \cite{wasserman1994social}, community detection has become a popular topic. It aimed to solve how to identify communities with groups of nodes, which have closer connection within the community and looser connection between the community. Detecting community has been proven to be helpful in loads of networks, i.e. biological networks, social media networks. It provides additional information about the structure and function of the whole network. Papadopoulos \cite{papadopoulos2012cd} once proposed a classification method of community detection algorithms based on their methodological principles and gave out the computational complexity as well as memory requirement of each referring algorithm. The vertex clustering method embeds graph nodes into a vector space to calculate distances between nodes. Spectral clustering, based on affinity matrix or Laplacian matrix, is a representative method in this subclass. Quality optimization can be the mainstream of community detection. The most rudimentary method GN algorithm was proposed by Girvan and Newman\cite{GN2002}, and then it was improved into a faster algorithm known as FN by Newman\cite{FN2004}. These greedy optimization methods intend to maximize modularity. Model-based methods include Label propagation\cite{Raghavan2007LPA} known as LPA and Infomap\cite{Rosvall2008Infomap}. LPA renews the label of nodes in the network through iteration and results in linear computation complexity, while Infomap introduces the theory of entropy of information.

Although there has been progress in the field of community detection, evidences show that overlapping community structure exists in most real networks \cite{palla2005uncovering,ahn2010link}, in which the overlapping nodes may have multiple significance. Several popular types of overlapping community detection methods have been proposed\cite{javed2018community}. Like maximizing modularity in non-overlapping community detection, the line of local expansion and optimization maximizes a local benefit function, which originally calculates the internal and external weights of community\cite{baumes2005finding}, and Lancichinetti\cite{lancichinetti2009detecting} proposed LFM method by using degree of communities instead of weights of communities. The series of Clique percolation method(CPM)\cite{palla2005uncovering} is based on the concept of K-clique, which refers to a complete subgraph with k nodes. There are several improvements in the CPM method \cite{kumpula2008sequential,maity2014extended}, but it is considered as a pattern recognizing algorithm rather than community detection\cite{xie2013overlapping}. As extensions of LPA, overlapping label propagation methods are proposed, including COPRA\cite{gregory2010finding}, SLPA\cite{xie2011slpa}, DEMON\cite{coscia2014uncovering} and NI-LPA\cite{el2020node}, which keep a fast linear time complexity. Besides, there are also methods based on the probabilistic graphical model and deep learning, i.e. BigCLAM\cite{yang2013overlapping} and CommunityGAN\cite{jia2019communitygan}. 

Networks' community structure has considerable influence on the propagation of information in networks. Some methods were invented in this way to improve the baseline ranking methods. Hu \cite{hu2013new} defined the KSC score as a weighted sum of internal and external scores, namely the k-shell score and community property. Zhao \cite{zhao2015community} proposed a novel stable centrality measure CbC(community-based centrality) to identify critically influential nodes. In order to decompose graphs separately using the k-shell algorithm, Luo \cite{luo2016identifying} differentiated weak ties from strong ties by community structure. However, in real networks, the observation that nodes are often clustering in one community could be problematic \cite{he2015novel}. CD-PR, an improvement of PageRank algorithm, extracted a certain proportion of nodes from each community according to the probability of community selection\cite{Wang2020Nodes}. But its community detection result is quite random, although it is fast to compute using LPA algorithm, which may lead to the instability of important nodes identification. Using the EnRenew algorithm based on information entropy, selected nodes will then be relatively dispersed in different communities and become more influential \cite{guo2020influential}. Combining local and global information, it refined the formula of VoteRank, but it used little information about the community itself, which might lead to unreliable results.

Recently, there have been methods to identify influential nodes with overlapping community structure. Some researches proposed immunization strategies to identify vital nodes, mostly based on membership locally, guaranteeing a low computation complexity. Taghavian\cite{taghavian2017local} ran a random-walk overlap selection(RWOS) to immunize the overlapping nodes in this process, and Kumar \cite{kumar2018efficient} chose to immunize highly connected neighbors of overlapping nodes in the network. But there is no theory to justify whether the extracted nodes by this process is important. Others considered both local and global information, including extending centrality measures \cite{ghalmane2019centrality}, defining overlapping modularity vitality to score each node\cite{rajeh2021identifying} and graph representation learning based methods\cite{wei2018identifying}. However, how to efficiently combine the results of local and global information using overlapping community still remains unresolved. In other words, it seems that existing methods did use the information of overlapping communities, but did not learn from the algorithms with non-overlapping community detection. Therefore, a compatible method to identify influential nodes using both overlapping and non-overlapping is anticipated.

There is a contradiction between the results' precision and algorithmic complexity of both non-overlapping and overlapping community detection methods. Sampling and resampling methods are applied here to decrease the uncertainty.
There are two main motivations for applying sampling in networks. One is that many networks' data are collected through 'crawlers', and these incomplete data could actually increase error in inference. The other is that most data mining methods are time-consuming, making them unsuitable for analyzing large scale networks. 
Metropolis Algorithms on the framework of Markov Chain Monte Carlo Methods is a classic graph sampling method \cite{hubler2008metropolis}. Besides, implicit characteristics such as structural properties and functional properties of the networks can also be inferred by sampling method \cite{maiya2011sampling}. Recently, the resampling method has been proposed. Inspired by bagging algorithm in machine learning \cite{breiman1996bagging}, Tixier newly proposed an algorithm by creating perturbed versions of the original graph and combining the results in order to decrease the variance of scoring when the result of base ranking algorithm is unstable such as graph degeneracy methods \cite{goltsev2006k}, which is named Perturb and Combined (P\&C) procedure \cite{tixier2019perturb}. Additionally, P\&C framework is proved to decrease the Mean Squared Error of node ranking.

In this paper, we propose a penalized method called RCD-Map experimented on five commonly used baseline ranking methods(Degree centrality, Closeness centrality, Betweennees centrality, K-shell and PageRank). The penalty is given based on local community structure as well as global network information. Resampling is used to achieve the comparably fast and accurate community detection result. Node ranking results are evaluated through SIR epidemic model. The experiment on real network datasets shows that our method RCD-Map with five baseline methods generally has a vaster propagation range than the original. 

Our main contributions are:
\begin{enumerate}
    \item Theoretically, RCD-Map, a new node ranking method with penalty of nodes' initial ranking scores for belonging to the same community is proposed, which is widely accessible to different baseline ranking methods with only a small adjustment on parameter. 
	\item By sampling and resampling, we fix the contradiction between results' precision and algorithmic complexity of both non-overlapping and overlapping community detection methods since community structure is better needed in the algorithm. Not only is the traditional non-overlapping community detection method applied here, but also overlapping community detection is used to test the universality of the algorithm.
	\item Use experiment on real networks of different sizes to prove that the proposed method RCD-Map is more efficient in information propagation.
\end{enumerate}

The remainder of this paper is organized as follows. Section 2 introduces the formulas of baseline scoring methods as well as the algorithms of community detection. In Section 3, details of the proposed method are introduced. Section 4 reports and analyzes the results of experiments, followed by discussions and a conclusion in Section 5.

\section{Preliminaries}
\subsection{Baseline methods}
We suppose the network is undirected and unweighted, denoted by $G(V,E)$, where $V$ and $E$ represent for the set of nodes and edges of network $G$. $n$ is defined as the number of nodes in network $G$.

In order to illustrate that our proposed method performs well in the field of influential nodes identification, we compare it with some baseline methods, including degree centrality \cite{bonacich1972factoring},closeness centrality \cite{freeman1978centrality}, betweenness centrality \cite{bavelas1948mathematical}, k-shell decomposition \cite{kitsak2010identification}, and PageRank \cite{brin1998anatomy}. Here we list the calculation of them.

\paragraph{Degree centrality (DC)} This method calculates the degree of each node as its score, i.e. $DC(i) = \sum_{j=1}^{n}a_{ij}$, where $a_{ij}$ represents if node $i$ is connected with node $j$. 

\paragraph{Closeness centrality (CC)} Closeness centrality calculates the reciprocal value of the sum of the shortest path length between one node and other nodes in the network. It is defined as $CC(i)=1/{\sum\limits_{j=1}^{n}d_{ij}}$, where $d_{ij}$ is the shortest path length between node $i$ to node $j$.  

\paragraph{Betweenness centrality (BC)} Betweenness centrality is based on the concept that the more shortest path pass through the node, the more important this node is. It describes nodes' control ability of network flow transported by the shortest path. It is proposed as $BC(i)=\sum\limits_{s,t\neq i}\dfrac{g_{st}(i)}{g_{st}}$, where $g_{st}$ is the number of shortest path from node $s$ to node $t$, and $g_{st}(i)$ means the number of path passing node $i$ among all $g_{st}$ paths.

\paragraph{K-shell decomposition} On the assumption that there is no isolated node in the network, the algorithm follows three steps. Step 1, from the perspective of degree centrality, the nodes with degree $DC=1$ are the least important nodes in the network, so all these nodes and their edges are removed. The remaining network will exist new nodes with degree $DC=1$ in some cases. This step will be continued until there exists no node with degree $DC=1$ in the network. Step 2, the nodes removed at the first step formed 1-core, and they are assigned with an importance index $ks=1$. Step 3, iterate the above two steps until all nodes are removed. In the $k^{\text{th}}$ iteration, the removed nodes formed k-core, and their importance is $ks=k$.

\paragraph{PageRank} The theoretical basis of PageRank is that one page's importance is determined by both the number and the quality of pages linked into it in the web. Every node's original PR index at time step $t$ is defined as $PR_{i}(t) = \sum_{j=1}^{n}a_{ji}\frac{PR_{i}(t-1)}{k_{j}^{out}}$, where $k_{j}^{out}$ is the out degree of node $v_{j}$. For fear that there may exist node with zero out degree, a skip probability c is set. The PR index is renewed as $PR_{i}(t) = (1-c)\sum_{j=1}^{n}a_{ji}\frac{PR_{i}(t-1)}{k_{j}^{out}} + \frac{c}{n}$. The PR index of one node is delivered equally to all the nodes in the network with the probability of c, and to the nodes into it with the probability of $1-c$. Studies showed that $c=0.15$ is a well performed parameter.

\subsection{Communtiy detection}
To make a trade-off between computation complexity and community detection's precision, we choose Infomap as non-overlapping community detection method, and DEMON is selected as overlapping community detection method. Besides, for the purpose of verifying the stability of different community detection methods, three alternative algorithms are added to each(non-overlapping:Infomap, GN and LPA, overlapping:DEMON,K-clique and BigCLAM) The basic ideas of the algorithms will be introduced below. 

\paragraph{Non-Overlapping community detection:} 

\paragraph{Infomap} Infomap uses the knowledge in information theory to find a map equation, which is the total minimum average coding length. What it is going to do is just find a way to minimize that index as small as possible. Without any hyperparameters, it moves the nodes in random sequential order, and each node is moved to the neighboring module that results in the largest decrease of the map equation.

\paragraph{GN} GN removes the edges with highest betweenness centrality, and recalculates the betweenness centrality of the remaining edges at each iteration. The algorithm ends with the biggest modularity value.

\paragraph{LPA} LPA gives each node an initial unique label, and at each iteration of the propagation, every node's label will be renewed as the most frequently appeared label of its neighbours belongs to. It converges when each node has the main label of its neighbours.

\paragraph{Overlapping community detection:} 

\paragraph{DEMON} DEMON is a local-first node-centrality-based approach to community discovery. In practice, the community structure of every ego network formed by each node and its neighborhood is discovered by applying the label propagation algorithm. One node's community will be determined by its peer neighbors. The overlapping communities will be collapsed as a single node and reconnected. The process continued until the entire network has been collapsed in a single community.

\paragraph{K-clique} K-clique first finds all complete subgraphs of size k, and a K-clique community is determined as the union of these subgraphs among which one shares k-1 nodes with another. 

\paragraph{BigCLAM} BigCLAM is based on the probabilistic graphical model, and it can be concluded as three steps: create a community-affiliation graph model, use gradient ascent to achieve graph fitting, and finally determine the community association.

\section{Proposed methods}
In this section, we will first explain the motivation of RCD-Map, then the pseudo-code of our framework, as well as details of the resampling and penalty methods, will be given. 

\subsection{Key idea}
\begin{figure}[!ht]
 	\centering
 	\caption{Top-5 important nodes selected by k-shell method}
 	\label{gathering}
 	\includegraphics[width=0.3\textheight]{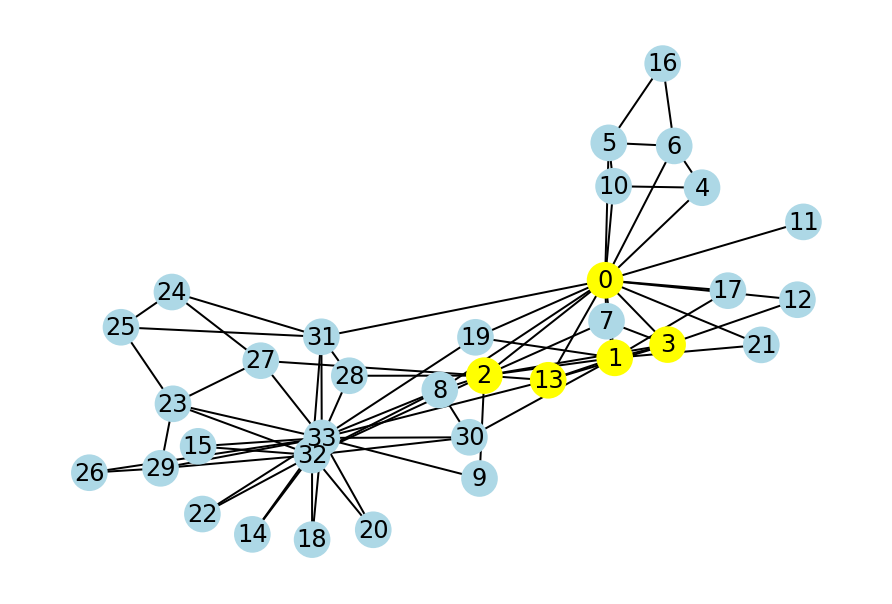}
\end{figure}

Information flow transports faster in the same community and more slowly between different communities\cite{hu2013new}, so it is preferable to choose nodes in different communities as initial nodes to achieve rapid and vast propagation. As Figure \ref{gathering} shows, however, gathering phenomena of important nodes exists in networks\cite{zhang2016identifying}, which means nodes of great importance are always in the same community. It hinders the efficiency of propagation in the network. Therefore, once one specific node has been chosen, we propose that the other nodes' importance in this community ought to be penalized. Especially, in overlapping community detection, nodes belonging to more than one community are named bridge nodes. They have a stronger influence on information propagation than others, so their penalties will be slighter. 

It is not easy to balance the computational complexity and accuracy in community detection. Since networks with factual community information are few, the most widely used accuracy remarking method is modularity Q. Most modularity optimization methods, i.e. GN and its improvements, are very costly in time with time-complexity $O(n^{3})$, so they are unsuitable for large size networks. Other model-based algorithms like LPA and Infomap has large randomness caused by different renewing order and labels chosen in the first iteration. The resampling theory is applied here to decrease the biases caused by community detection. 

Only applying non-overlapping community detection algorithms is less proper since overlapping community structure exists in most real networks. Besides, the penalty of these two algorithms should not be the same, otherwise, the scores of important overlapping nodes will decrease more. Therefore, we propose a compatible algorithm suitable to both community detection theories.

\begin{algorithm}[ht]
    \caption{Pseudo-code of RCD-Map framework}
    \label{algorithm}
    \renewcommand{\algorithmicrequire}{\textbf{Input:}}
    \renewcommand{\algorithmicensure}{\textbf{Output:}}
    \renewcommand{\algorithmiccomment}[1]{//#1}
    \begin{algorithmic}[1]
    \REQUIRE Graph $G(V,E)$ with $V=\{v_1,\cdots,v_n\}$ and $E=\{e_1,\cdots,e_m\}$, a base vertex scoring function $s:V\rightarrow R$, community detection algorithm $f$, and $M$ denotes the resampling times
    \ENSURE S contains nodes and their scores in order of influence
    \FOR{$i = 1; i\le M;i++$}
        \STATE Handle RCD$(G,f)$ Algorithm \ref{algorithm2};
        \STATE Get Base Ranking result $\text{initres} = s(V_i)$ for all nodes;
        \STATE Handle Map Algorithm \ref{algorithm3};
    \ENDFOR
    \STATE Averaging $M$ results to obtain the final ranking result
    \end{algorithmic}
\end{algorithm}
As the pseudo-code of our framework shows, RCD-Map(\textbf{R}esampling \textbf{C}ommunity \textbf{D}etection to \textbf{Ma}ximize \textbf{p}ropagation in complex networks) is a meta-algorithm that mainly contains two parts: resampling to detect community and using community information as penalty to maximize propagation, which include detailed links Algorithm\ref{algorithm2},\ref{algorithm3}. We will introduce them in the following parts.

\subsection{Resampling to detect community}
Three types of perturbation noise model have been proposed, Uniform Perturbation(ERP), Degree Assortative Perturbation(CLP), and Link Prediction Based Model(LPP). ERP is chosen as our resampling method since previous research showed that top ranking nodes seem stable in the ERP model \cite{Adiga2013HowRI}. In the following part, we will explicate the resampling framework and then the details of ERP.

\paragraph{Resampling framework}
The resampling framework is similar to the Uniform perturbation noise model. 
This edge-based perturbation model is originally used to test the robustness of core decomposition. The formula of the model $\varTheta(G,G_{ER},\epsilon)$ is defined as: 
\[ P_{\varTheta}[(u,v)] =
	\epsilon P_{G_{ER}}[(u,v)],
\]
\noindent where $(u,v)$ is the existing edge of $G_{ER}$, $P_{\varTheta}[(u,v)]$ is the probability of adding or deleting the edge between node $i$ and node $j$, $G_{ER}$ is the corresponding Uniform perturbation noise model of the original graph $G$, $P_{G_{ER}}[(u,v)]$ is the possibility of choosing edge $(u,v)$ in $G_{ER}$, and $\epsilon$ is the probability of changing one edge. 
Once $G$ is given, the ERP of it can be modeled. At each run, every existed or potential connection $(u,v)$ of $G$ is judged whether its connection condition will be altered. The possibility of perturbation is $P_{\varTheta}[(u,v)]$ as given above. Then, for every connection, a number from uniform distribution $U(0,1)$ is generated. If this number is smaller than the perturbation possibility, the connection condition will be changed, i.e. there will be a new edge if two nodes are not connected originally and vice versa. Details of the algorithm can be found in Algorithm \ref{algorithm2}.

\begin{algorithm}[ht]
    \caption{Pseudo-code of RCD(Resampling Community Detection)}
    \label{algorithm2}
    \renewcommand{\algorithmicrequire}{\textbf{Input:}}
    \renewcommand{\algorithmicensure}{\textbf{Output:}}
    \renewcommand{\algorithmiccomment}[1]{//#1}
    \begin{algorithmic}[1]
    \REQUIRE Graph $G(V,E)$ with $V=\{v_1,\cdots,v_n\}$ and $E=\{e_1,\cdots,e_m\}$, community detection algorithm $f$
    \ENSURE the community detection result $G_i^{*}$
    \STATE \COMMENT{Resampling the edges to perturb the graph}
    \FORALL{$e_{ij}$ in $E_{ER}$}
    \STATE $\epsilon \propto \frac{1}{|E_{ER}|}$
        \IF{$U(0,1) < \epsilon $}
        \STATE $\text{Change\_edge}(e_{ij})$
        \ENDIF
    \ENDFOR
    \STATE $G_i^* = f(G_i)$ \COMMENT{Community Detection}
    \end{algorithmic}
\end{algorithm}

\paragraph{Uniform perturbation (ERP)}
This model uses the concept of ER random model proposed by Erd{\H{o}}s and R{\'e}nyi\cite{erdHos1960evolution}. The resampling source model is set as $G_{ER}(n,\frac{1}{n})$, which means that the nodes in $G_{ER}$ are the same as those in $G$, and the connection probability between two nodes is $\frac{1}{n}$. In resampling, every edge is chosen with the same probability $P_{G_{ER}}[(u,v)]=\frac{1}{\|V_{ER}\|}$. 

\subsection{Penalized iterative nodes selection}

According to the key ideas above, for the purpose of getting more influential propagation ranking results than the baseline ranking algorithm, after each run of resampling to perturb the original graph, we iteratively select influential nodes with rescoring by community penalty. The penalty for other nodes in the same community with the selected node, i.e. for node $u$ if node $v$ has been chosen, is defined as:

\[
s'(u) = s(u)-\frac{\alpha}{n_{C(i)}\|C_{v}\|^k},
\]

\noindent where $s'(u)$ is node $u$'s score after penalizing, $s(u)$ is the original form, $n_{C(i)}$ is the total number of communities including node $i$ in overlapping community detection, and $\|C_{v}\|$ denotes the number of nodes in the community $C_{v}$.

In non-overlapping community detection, $k$ is set as $1$. In overlapping community structure, since it is believed that bridge nodes have a stronger influence, they should be given priority to choose, and the penalties of them will be slighter. So, we reduce the order of magnitude in community penalty by exponential parameter $k$. Empirically, we set $k=2$.

\begin{algorithm}[ht]
    \caption{Pseudo-code of Map(Maximize propagation)}
    \label{algorithm3}
    \renewcommand{\algorithmicrequire}{\textbf{Input:}}
    \renewcommand{\algorithmicensure}{\textbf{Output:}}
    \renewcommand{\algorithmiccomment}[1]{//#1}
    \begin{algorithmic}[1]
    \REQUIRE baseline ranking result without community information $s(v)$, and the community detection result $G_i^{*}$
    \ENSURE $s$ contains nodes and their scores in this iteration
    \WHILE{$|V_i^*|>0$}
        \STATE add $v$ to $S_i$, where $v = \mathop{\arg\max}\limits_{v \in V_i^*} s(v)$
        \STATE $V_i^*\leftarrow V_i^* \setminus \{v\}$
        \FORALL{node $u$ in $v$'s community $C_v$}
        \STATE $s(u) -= \alpha \times \frac{1}{\text{len}(C_v)}$
        \ENDFOR
    \ENDWHILE
    \end{algorithmic}
\end{algorithm}

Referring to VoteRank, where every node's initial voting ability is 1, then node's initial score is the sum of its neighbors' voting ability, namely its degree. The coefficient setting in VoteRank is the inverse of average degree of the network. In RCD-Map, every node's initial score is marked by baseline ranking method, so we support that the coefficient $\alpha$ should be varied based on the ranking method. For degree centrality, closeness centrality, and betweenness centrality, $\alpha$ is set as the inverse of average degree, average shortest path length, and average beweenness of the network. For k-shell, since there exists limitations that it leaves some nodes with the same ranking and the gap between different score is at least 1, $\alpha$ should be set as a bigger number to ensure that penalty is useful. We set the coefficient as $max(ks)/2.5$. For PageRank, in contrast, the score is the skip probability with maximum value even less than one percent in large networks. $\alpha$ is set as the maximum value of PageRank score instead. The detailed algorithm is presented in Algorithm \ref{algorithm3}.

\section{Results}
In this section, we will first introduce the topological characteristics of selected networks, and then the Karate network will be used as an example to explain why RCD-Map has decided advantage. This superiority will be verified more precisely through SIR model to see the information propagation ability of RCD-Map. Besides, synthetic networks with known communities are also tested in this part.

\subsection{Data description}
Since our method is proposed based on the characteristics of real networks, we experimented it on four well-known real social networks. These networks used are undirected and unweighted, and we only consider the problem in their maximum connected components.

\begin{enumerate}
    \item \emph{Karate Club} is a classical social network with a clear community structure. Zachary captures 34 members of a karate club, documenting links between pairs of members who interacted outside the club. \cite{zachary1977information}
    \item \emph{Jazz Musicians} is a collaboration social network in which each node denotes a musician and each edge represents collaboration in a band. \cite{gleiser2003community}
    \item \emph{Email} is an email communication network, where nodes are users and each edge represents that at least one email was sent. \cite{guimera2003self}
    \item \emph{Wiki Vote} is an election social network. Nodes in the network are Wikipedia users and an edge between two nodes means that one have voted the other. \cite{leskovec2010signed}
\end{enumerate}

\begin{table}[!ht]
\centering
\caption{The topological characteristics of four real networks}
\label{tab:datasets}
\begin{tabular}{cccccc}
\hline
$G(V,E)$       & $|V|$ & $|E|$ & $\langle k \rangle$ & $k_{max}$ & $\tau$ \\ \hline
Karate Club    & 34    & 78    & 4.59  & 17        & 6.73   \\
Jazz Musicians & 198   & 2742  & 27.70 & 100       & 40.03  \\
Email          & 1133  & 5452  & 9.62  & 71        & 20.75  \\ 
Wiki Vote & 7066 & 103663 & 6 & 102 & 138.15 \\ \hline
\end{tabular}
\end{table}

The statistical properties are shown in Table \ref{tab:datasets}, where $\langle k \rangle$ and $k_{max}$ denote the average and maximum degree in the graph, and $\tau$ is the maximum eigenvalues of the adjacency matrix.

\subsection{Experimental details of an example network}

\begin{figure}[!ht]
    		\centering
    		\caption{Illustration of node selection before and after adopting the new method}
    		\begin{minipage}[c]{0.45\textwidth}
    			\centering
    			\subcaption{Top-1 node by closeness method}
    			\includegraphics[height=0.2\textheight]{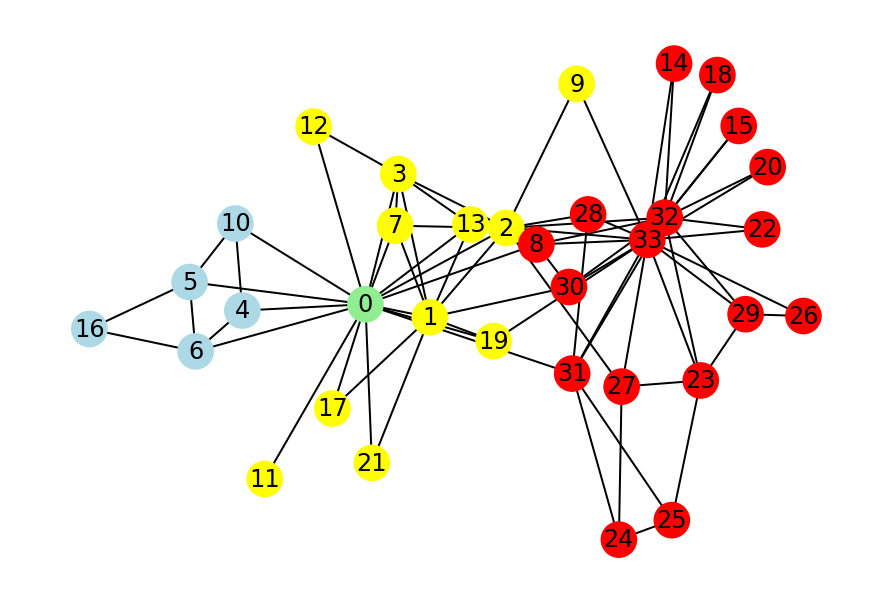}
    			\label{fig:Karate-closeness-base-1}
    		\end{minipage}
    		\begin{minipage}[c]{0.45\textwidth}
    			\centering
    			\subcaption{Top-2 nodes by closeness method}
    			\includegraphics[height=0.2\textheight]{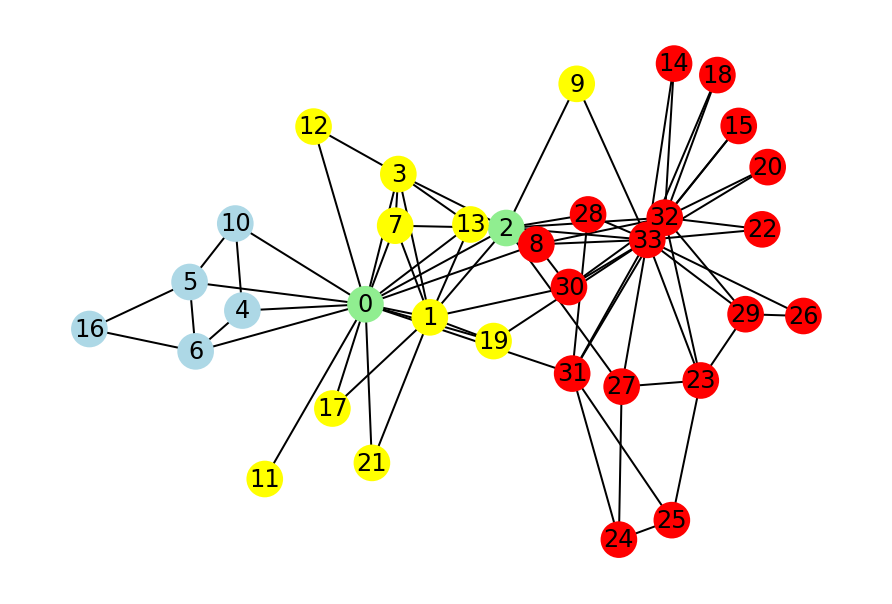}
    			\label{fig:Karate-closeness-base-2}
    		\end{minipage}
    		\begin{minipage}[c]{0.45\textwidth}
    			\centering
    			\subcaption{Top-1 node by RCD-Map}
    			\includegraphics[height=0.2\textheight]{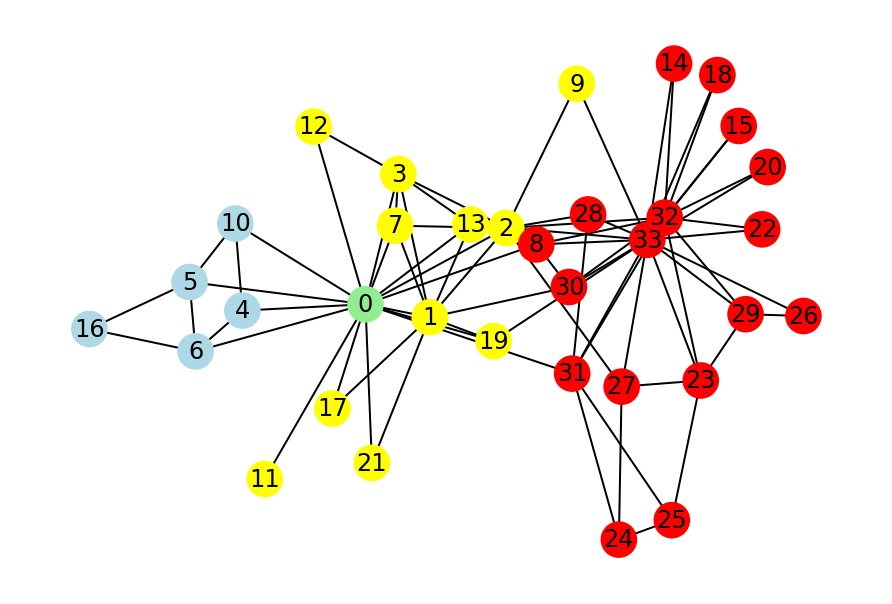}
    			\label{fig:Karate-closeness-RCD-Map-1}
    		\end{minipage}
    		\begin{minipage}[c]{0.45\textwidth}
    			\centering
    			\subcaption{Top-2 nodes by RCD-Map}
    			\includegraphics[height=0.2\textheight]{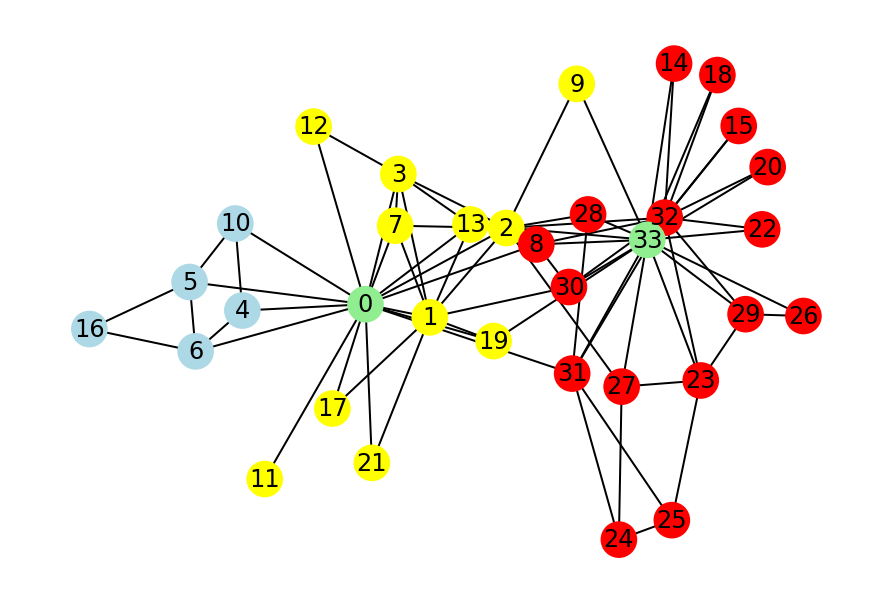}
    			\label{fig:Karate-closeness-RCD-Map-2}
    		\end{minipage}
    		\label{fig:Karate-detail}
\end{figure}

We illustrate the experimental details of the Karate Club dataset below to show the rationality of nodes the proposed algorithm chooses. In Figure \ref{fig:Karate-detail}, the original community information is labeled by three different colors, i.e., red, yellow, and blue. Compared with baseline closeness centrality, in the corresponding RCD-Map result the top-2 nodes are distributed in different communities, which would have a broader spreading effect. This is because after picking node 0, we penalized the scores in the yellow community, then node 33 in the red community is followed by it to be selected.

Generally, as Table \ref{tab:example-Karate} shows, our proposed method can select top nodes in different communities. Before RCD-Map, all baseline methods do not pick any nodes from the first community into top five nodes. In following parts, we will show that through our dispersing method it improves the spreading ranges of information flow in simulations of propagation dynamics.

\begin{table}[!ht]
\centering
\caption{Experimental results on Karate Club dataset}
\label{tab:example-Karate}
\begin{tabular}{ccccccccccccccccccc}
\hline
\multirow{2}{*}{Method} & \multirow{2}{*}{Algorithm} & \multicolumn{6}{c}{Initially selected nodes} & \multicolumn{11}{c}{Community Distribution}  \\ \cline{3-16} 
            &               & 1  & 2  & 3  & 4  & 5  & 6  & & & 1 & & 2 & & 3 & & & \\ \hline
Centrality  & Base          & 33 & 0  & 32 & 2  & 1  & 3  & & & 0 & & 4 & & 2 & & & \\
            & RCD-Map & 33 & 0  & 32 & 2  & 1  & 31 & & & 0 & & 3 & & 3 & & & \\
Closeness   & Base          & 0  & 2  & 33 & 31 & 8  & 13 & & & 0 & & 3 & & 3 & & & \\
            & RCD-Map & 0  & 33 & 2  & 31 & 8  & 13 & & & 0 & & 3 & & 3 & & & \\
Betweenness & Base          & 0  & 33 & 32 & 2  & 31 & 8  & & & 0 & & 2 & & 4 & & & \\
            & RCD-Map & 0  & 33 & 32 & 2  & 31 & 5  & & & 1 & & 2 & & 3 & & & \\
K-shell     & Base          & 0  & 1  & 2  & 3  & 13 & 7  & & & 0 & & 6 & & 0 & & & \\
            & RCD-Map & 0  & 8  & 1  & 30 & 4  & 2  & & & 1 & & 3 & & 2 & & & \\
PageRank    & Base          & 33 & 0  & 32 & 2  & 1  & 31 & & & 0 & & 3 & & 3 & & & \\
            & RCD-Map & 33 & 0  & 5  & 32 & 2  & 31 & & & 1 & & 2 & & 3 & & & \\ \hline
\end{tabular}
\end{table}

\subsection{Simulations of propagation dynamics}

To evaluate the selected lists, we simulate the spreading of SIR(susceptible-infected-recovered) epidemic model\cite{kitsak2010identification}, which is a popular model of epidemic dynamics on complex networks, often used to illustrate the spreading of information. At each step of the SIR model, every node belongs to one of three statuses: Suspective(S), Infected(I), and Removed(R). Especially, in the original epidemic condition, 'R' represents the sum of recovered and dead people. Here we select the influential nodes and set them as initial spreaders, then observe how many nodes have been infected finally to assess the algorithm.

\begin{figure}[!ht]
 	\centering
 	\caption{States and flows in the SIR model}
 	\label{SIR-flowchart}
 	\includegraphics[width=0.3\textheight]{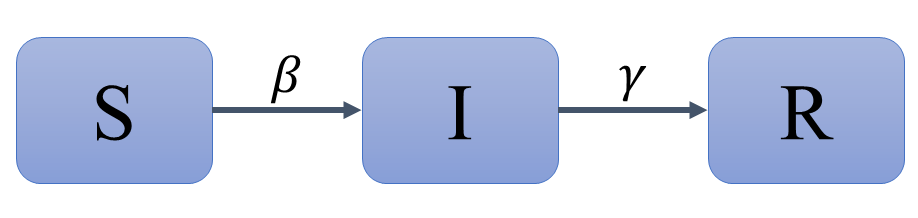}
\end{figure}

As Figure \ref{SIR-flowchart} shows, suppose at the $t$ step, the amounts of S, I, R are respectively $s(t)$, $i(t)$, and $r(t)$. In the beginning, all the nodes are suspective except that the top ranking nodes are infected. Then in each time step, $\beta$ and $\gamma$ denote the infected rate of S-nodes and removed rate of I-nodes. Differential equation theory obtains that $\sigma = \frac{\beta}{\gamma}<1$ is the necessary and sufficient condition to control the disease\cite{levin2012applied}, i.e. let the number of statuses converges, namely $s(t_c)$, $i(t_c)$, and $r(t_c)$. In our simulation we set $\beta=1/\tau$, where $\tau$ is the maximum eigenvalue of the adjacency matrix, and $\gamma=0.8$ \cite{chakrabarti2008epidemic}, because a larger $\sigma$ may lead to infecting all the nodes. Since there exists randomness in the spreading of SIR model, we average the number of nodes that has been infected after 1000 independent runs. 

\paragraph{Different numbers of initial nodes}
In our method, there are several parameters to set, including the resampling probability, community detection methods (both overlapping and non-overlapping), and the selection of $\alpha$.
\begin{figure}[!ht]
 	\centering
 	\caption{Relationship between the number of initial nodes and finally infected nodes}
 	\label{fig:initial}
 	\includegraphics[width=0.5\textheight]{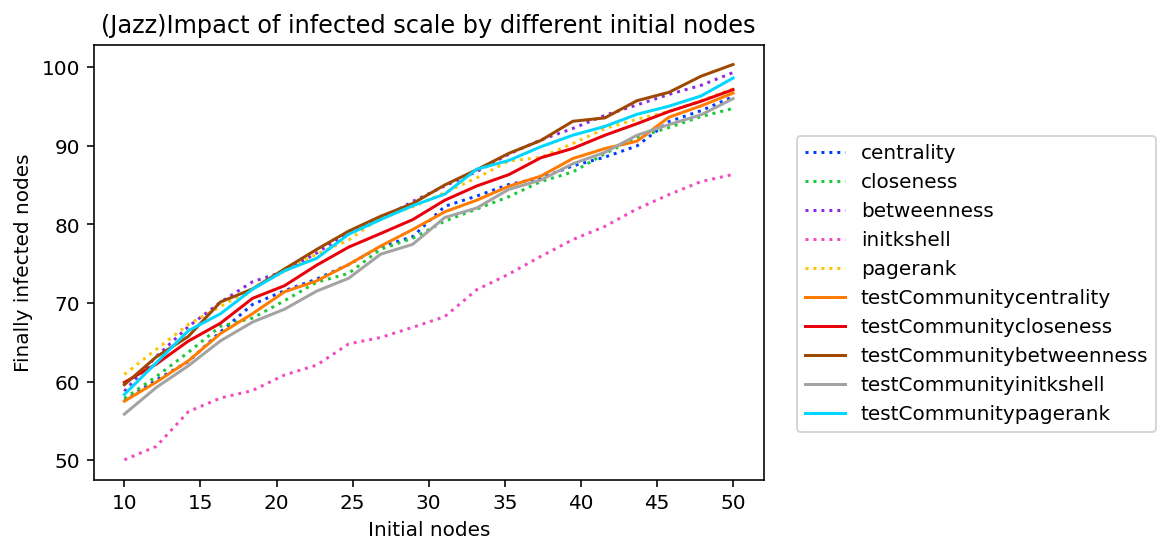}
\end{figure}

Figure \ref{fig:initial} shows the final simulation results when different numbers of the top-ranked nodes are chosen as infected nodes and Infomap is used as the community detection method in the Jazz-musician network. In general, RCD-Map improved in each baseline ranking method performs better than the original baseline ranking method with moderate numbers of initial nodes. Specifically, there is a remarkable increase in the propagation performance of RCD-Map improved in k-shell compared with the original k-shell. This figure indicates that the proposed method improved in most ranking methods has a larger propagation range when the number of initially infected nodes is medium, neither too large to be distinguished nor too small to spread information. 

\begin{table}[!ht]
\centering
\caption{Number of finally infected nodes in four datasets}
\label{tab:SIR-res-1}
\begin{tabular}{ccccccc}
\hline
       &                          & Centrality & Closeness & Betweenness & K-shell & PageRank \\ \hline
Karate & Base                     & 17.996     & 17.58     & 18.308      & 17.2565 & 17.987   \\
       & NonOverlap-RCD-Map & 18.75     & \textcolor{red}{18.584}    & 18.732      & \textcolor{red}{18.66}  & \textcolor{red}{19.123}   \\
       & Overlap-RCD-Map    & 18.288     & 17.704    & 18.448      & 18.043  & \textcolor{blue}{19.322}  \\
Jazz   & Base                     & 85.2275    & 84.5925   & 89.652      & 74.836  & 87.947   \\
       & NonOverlap-RCD-Map & 85.481     & \textcolor{red}{87.063}    & 90.091       & \textcolor{red}{81.974}  & 87.718   \\
       & Overlap-RCD-Map   & 85.407     & \textcolor{blue}{88.436}    & 90.225      & 80.128  & 88.266   \\
Email  & Base                     & 323.267   & 314.326  & 335.318      & 294.29 & 330.677 \\
       & NonOverlap-RCD-Map & \textcolor{red}{334.184}     & \textcolor{red}{334.184}   & 334.184    & \textcolor{red}{331.328} & 336.165  \\
       & Overlap-RCD-Map    & \textcolor{blue}{349.029}    & 335.692   & 335.447   & 302.248 & \textcolor{blue}{355.677}  \\ 
Wiki  & Base                     & 840.934   & 821.177  & 848.461      & 798.736 & 845.437 \\
       & NonOverlap-RCD-Map & 841.655     & \textcolor{red}{828.999}   & 849.241   & \textcolor{red}{825.229} & 846.822  \\
       & Overlap-RCD-Map    & 841.798    & \textcolor{blue}{835.684}   & 850.545         & 803.56 & \textcolor{blue}{854.009}  \\ \hline
\end{tabular}
\end{table}

\begin{figure}[!ht]
    		\centering
    		\caption{Averaging SIR spreading results by non-overlapping RCD-Map on four datasets}
    		\begin{minipage}[c]{0.45\textwidth}
    			\centering
    			\includegraphics[height=0.225\textheight]{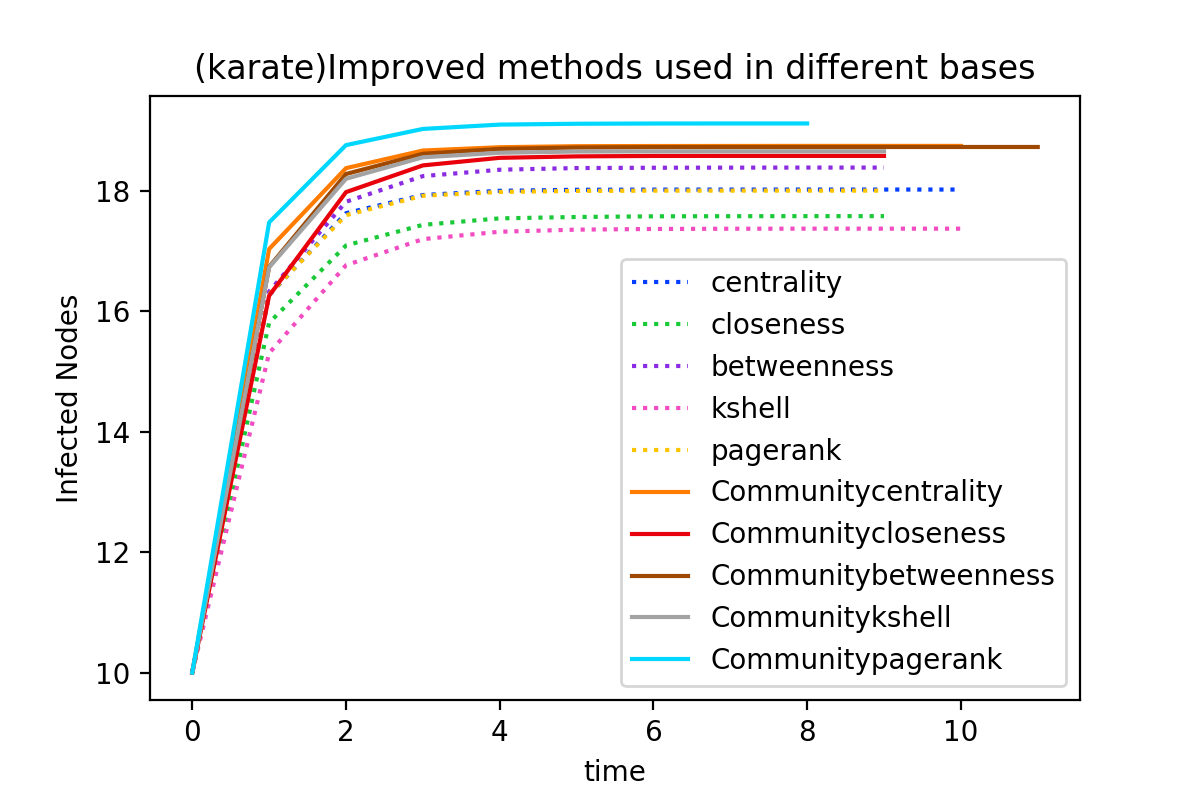}
    			\subcaption{SIR-nonoverlap-Karate}
    			\label{fig:SIR-nonoverlap-Karate}
    		\end{minipage}
    		\begin{minipage}[c]{0.45\textwidth}
    			\centering
    			\includegraphics[height=0.225\textheight]{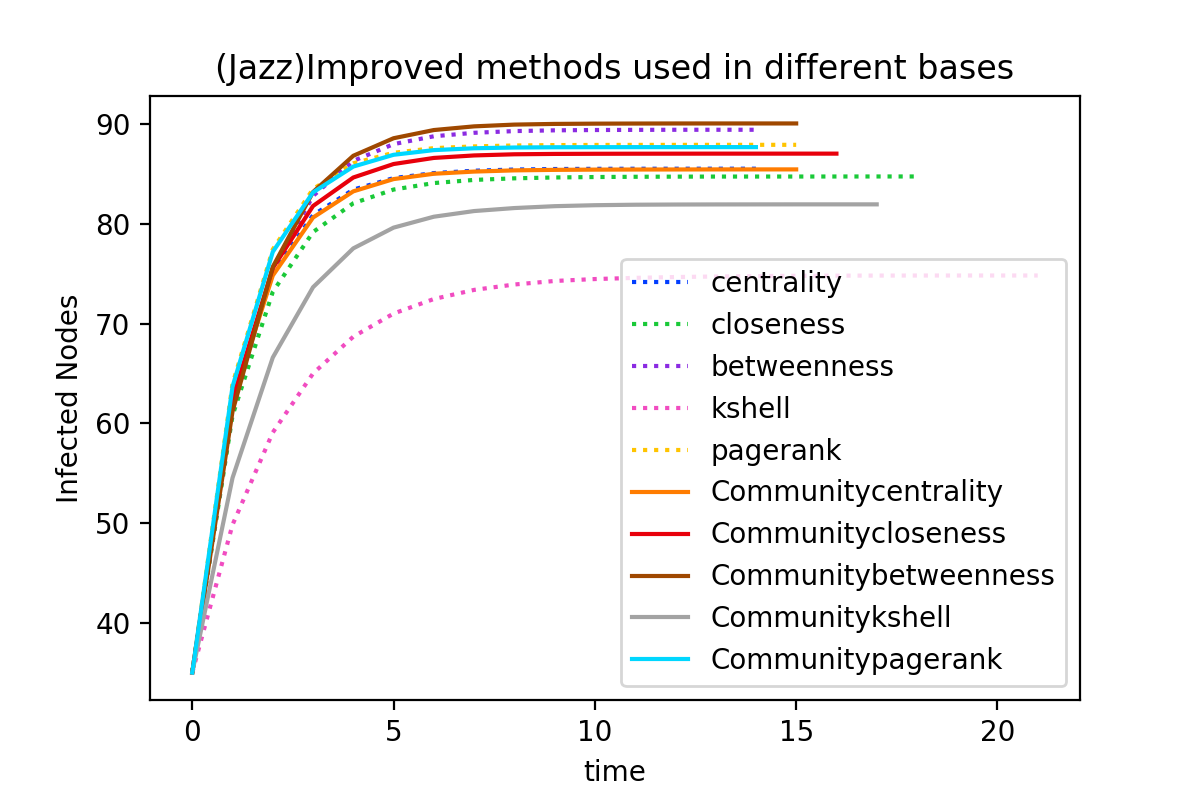}
    			\subcaption{SIR-nonoverlap-Jazz}
    			\label{fig:SIR-nonoverlap-Jazz}
    		\end{minipage}
    		\begin{minipage}[c]{0.45\textwidth}
    			\centering
    			\includegraphics[height=0.225\textheight]{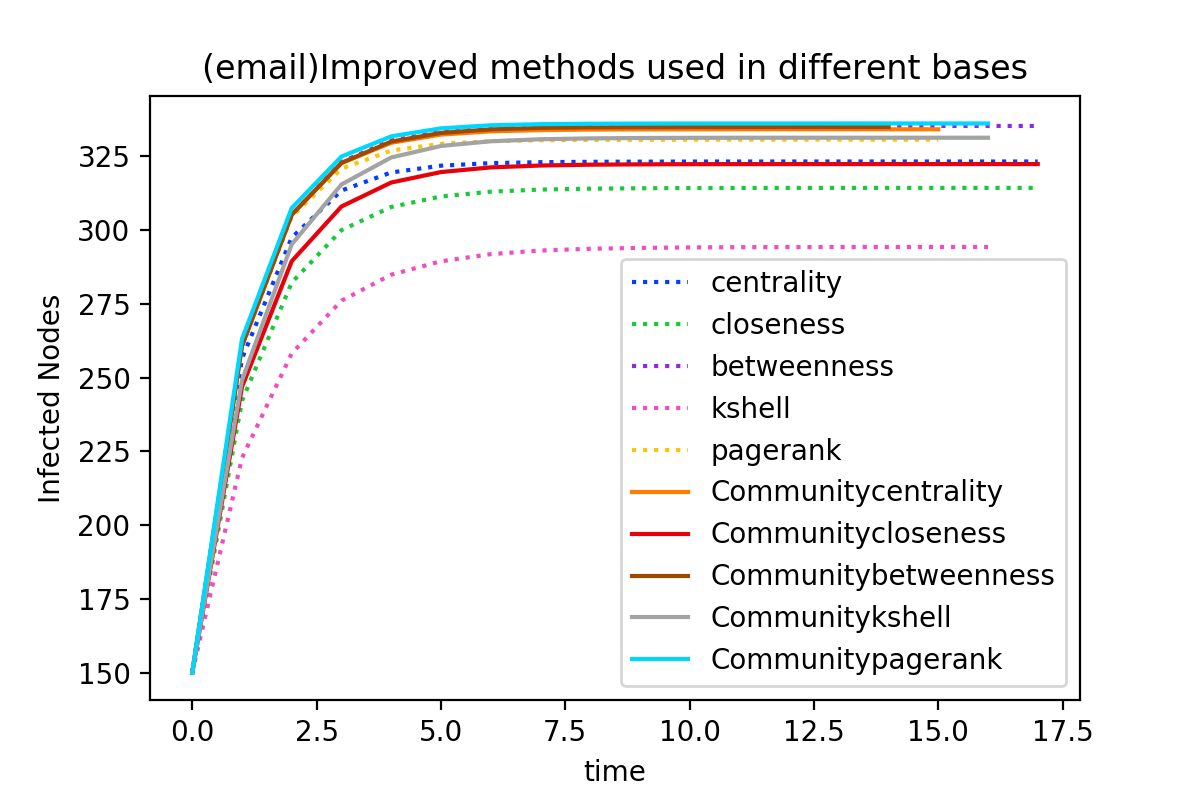}
    			\subcaption{SIR-nonoverlap-Email}
    			\label{fig:SIR-nonoverlap-Email}
    		\end{minipage}
    		\begin{minipage}[c]{0.45\textwidth}
    			\centering
    			\includegraphics[height=0.225\textheight]{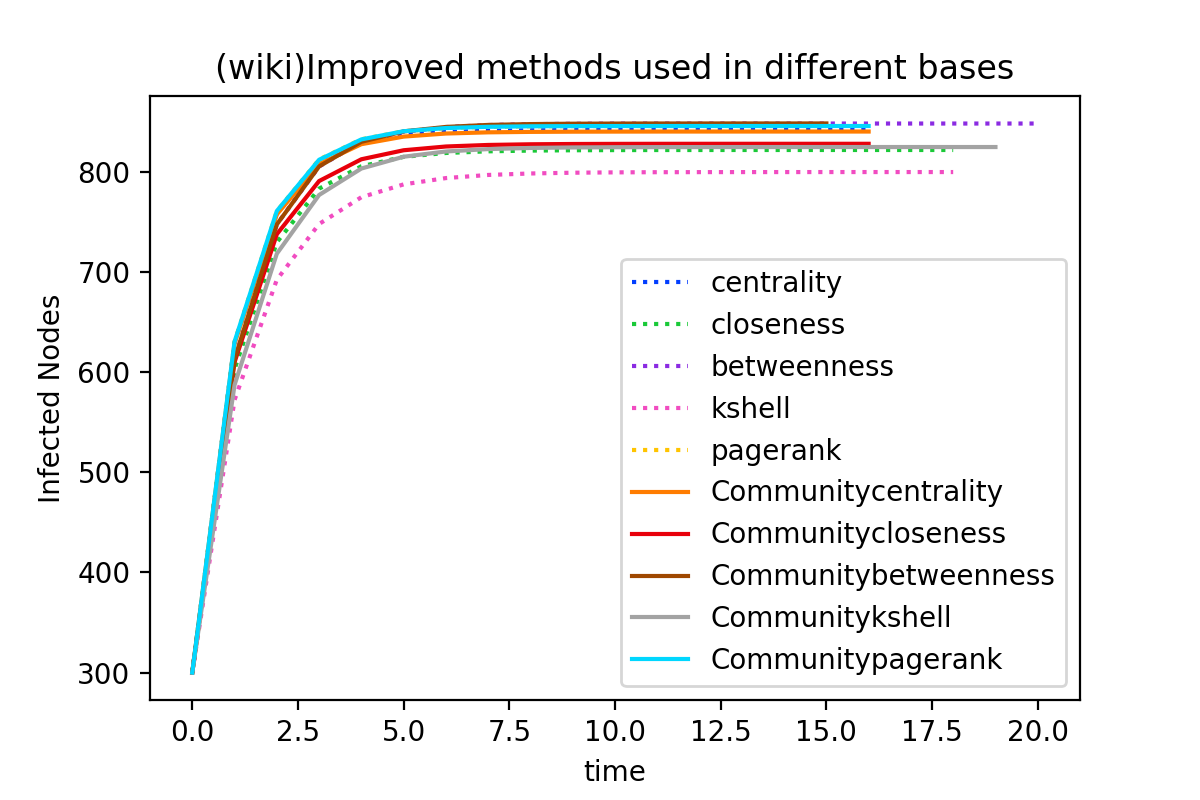}
    			\subcaption{SIR-nonoverlap-Wiki}
    			\label{fig:SIR-nonoverlap-Wiki}
    		\end{minipage}
    		\label{fig:SIR-nonoverlap}
\end{figure}

\begin{figure}[!ht]
    		\centering
    		\caption{Averaging SIR spreading results by overlapping RCD-Map on four datasets}
    		\begin{minipage}[c]{0.45\textwidth}
    			\centering
    			\includegraphics[height=0.225\textheight]{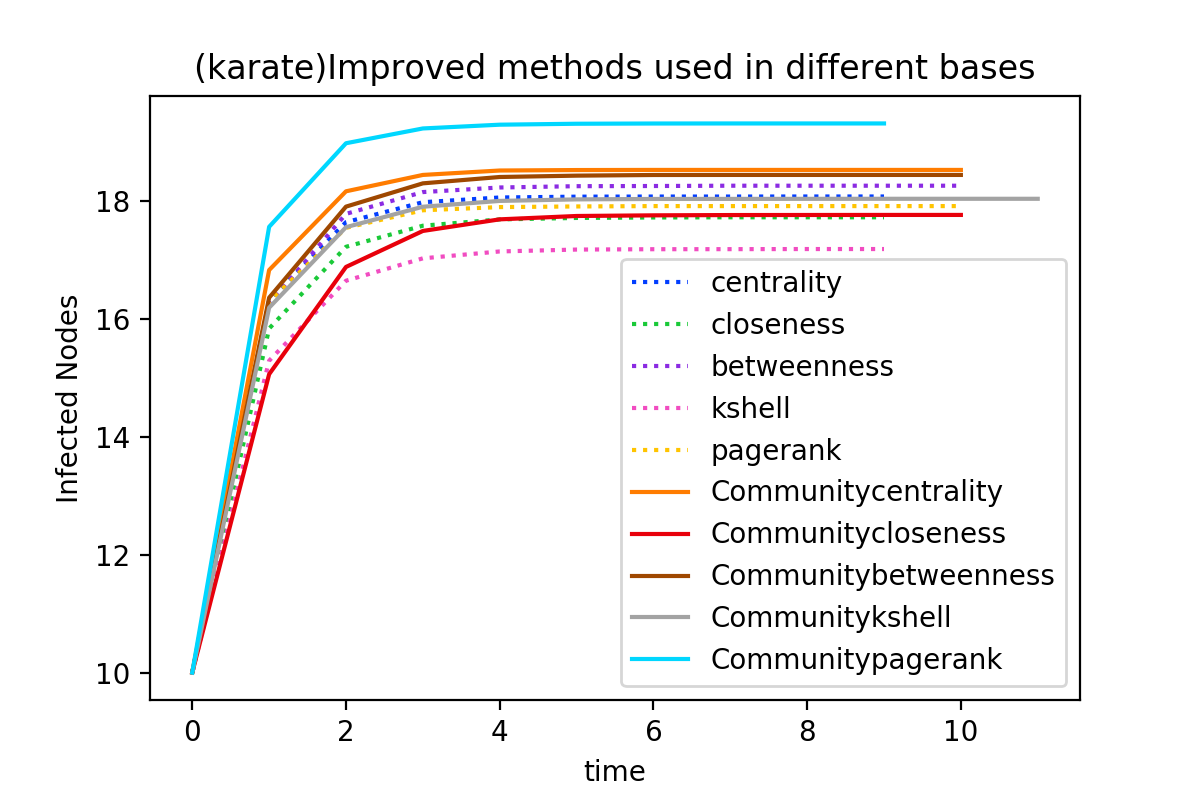}
    			\subcaption{SIR-overlap-Karate}
    			\label{fig:SIR-overlap-Karate}
    		\end{minipage}
    		\begin{minipage}[c]{0.45\textwidth}
    			\centering
    			\includegraphics[height=0.225\textheight]{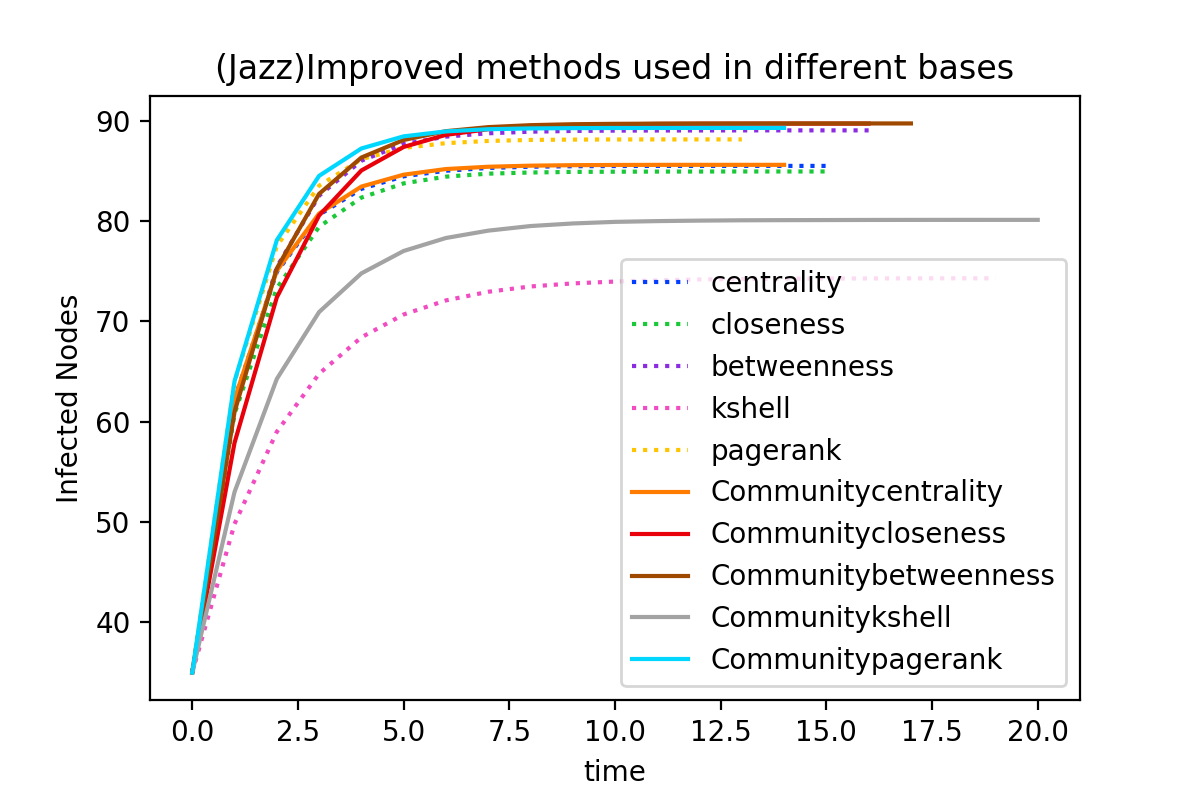}
    			\subcaption{SIR-overlap-Jazz}
    			\label{fig:SIR-overlap-Jazz}
    		\end{minipage}
    		\begin{minipage}[c]{0.45\textwidth}
    			\centering
    			\includegraphics[height=0.225\textheight]{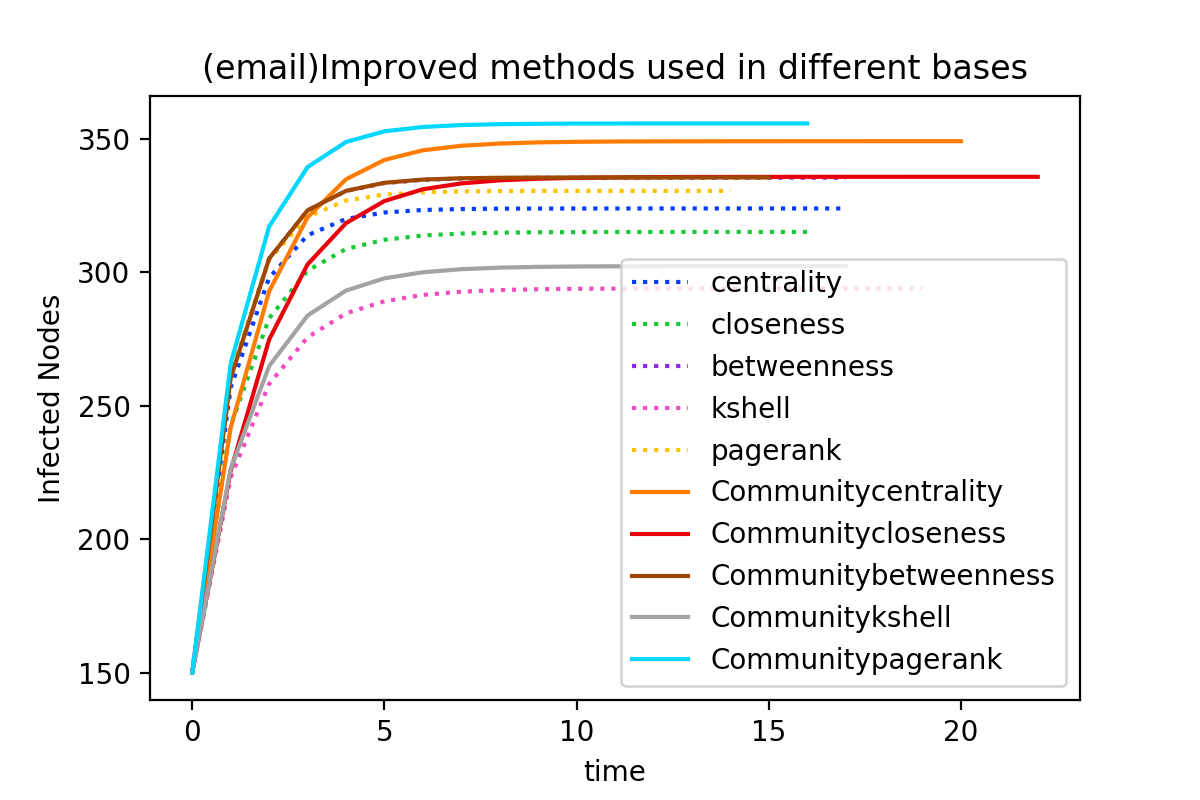}
    			\subcaption{SIR-overlap-Email}
    			\label{fig:SIR-overlap-Email}
    		\end{minipage}
    		\begin{minipage}[c]{0.45\textwidth}
    			\centering
    			\includegraphics[height=0.225\textheight]{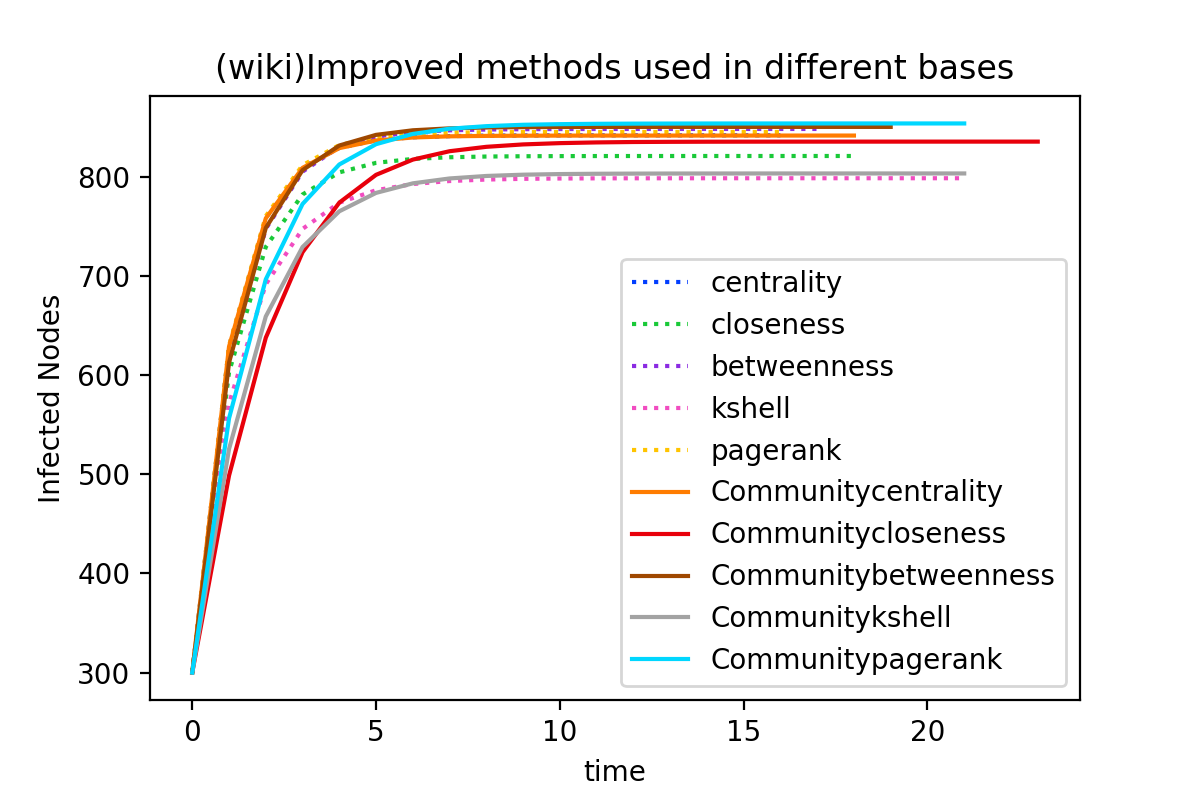}
    			\subcaption{SIR-overlap-Wiki}
    			\label{fig:SIR-overlap-Wiki}
    		\end{minipage}
    		\label{fig:SIR-overlap}
\end{figure}

\paragraph{Comparing the spreading ability of top-ranked nodes} Table \ref{tab:SIR-res-1} illustrates the average number of finally infected nodes in all four datasets. Figure \ref{fig:SIR-nonoverlap} and Figure \ref{fig:SIR-overlap} provide a visual presentation of our experimental results. In general, RCD-Map improved in five baseline methods DC,CC,BC,k-shell and PageRank, all achieving better spreading performance than their original forms separately. 

Figure \ref{fig:SIR-nonoverlap} shows that, in the experiment based on non-overlapping community detection, RCD-Map used on k-shell has remarkable promotes in all four datasets, especially in the larger-sized networks like Email and Wiki, where the final spreading range is 30 nodes more than the origin respectively. That is because k-shell is a coarse-grained ranking method. Besides, the spreading results of top-ranking nodes using RCD-Map improved in baseline methods DC,CC,and PageRank also see great increases to some extent. CC improves a lot in the Karate network as well as in the Jazz musician network, DC experiences a remarkable increase in the Email network, and PageRank increase sharply in the Karate network. Overall, using non-overlapping community detection, RCD-Map improved in baseline methods achieves better propagation performance with faster speed and larger range.

Figure \ref{fig:SIR-overlap} depicts that, in general, RCD-Map enhanced in different methods can achieve a vaster propagation range. Specifically, RCD-Map improved in DC and PageRank have decided advantage than original ones in the Email network. Their performance is even better than the counterpart using non-overlapping community detection. RCD-Map improved in DC and PageRank see more than 20 final infected nodes in the Email network separately. CC enhanced by RCD-Map experiences a boost in results in the Jazz network and the Wiki network. Despite the fact that the spreading rate of RCD-Map used on CC is somehow slower, the final spreading range is much vaster. To sum up, using overlapping community detection, baseline methods improved by RCD-Map can have a similar or stronger spreading ability. 

\paragraph{Synthetic networks with known community information}
In this part, LFR synthetic benchmark has been introduced to create networks with prior known communities\cite{Lancichinetti_2008}. The procedure follows four steps: first assign a power law distribution with parameter $\tau_1$ the degree sequence will follow, then generate another power law distribution with parameter $\tau_2$ the community sizes will follow, next each node will be given a community, and finally $(1 - \mu) \mathrm{degree}(u)$ numbers of intra-community edges as well as $\mu \mathrm{degree}(u)$ numbers of inter-community edges will be added to nodes $u$. 

\begin{figure}[!ht]
    		\centering
    		\caption{Averaging SIR spreading results with given community information on synthetic networks}
    		\begin{minipage}[c]{0.45\textwidth}
    			\centering
    			\includegraphics[height=0.225\textheight]{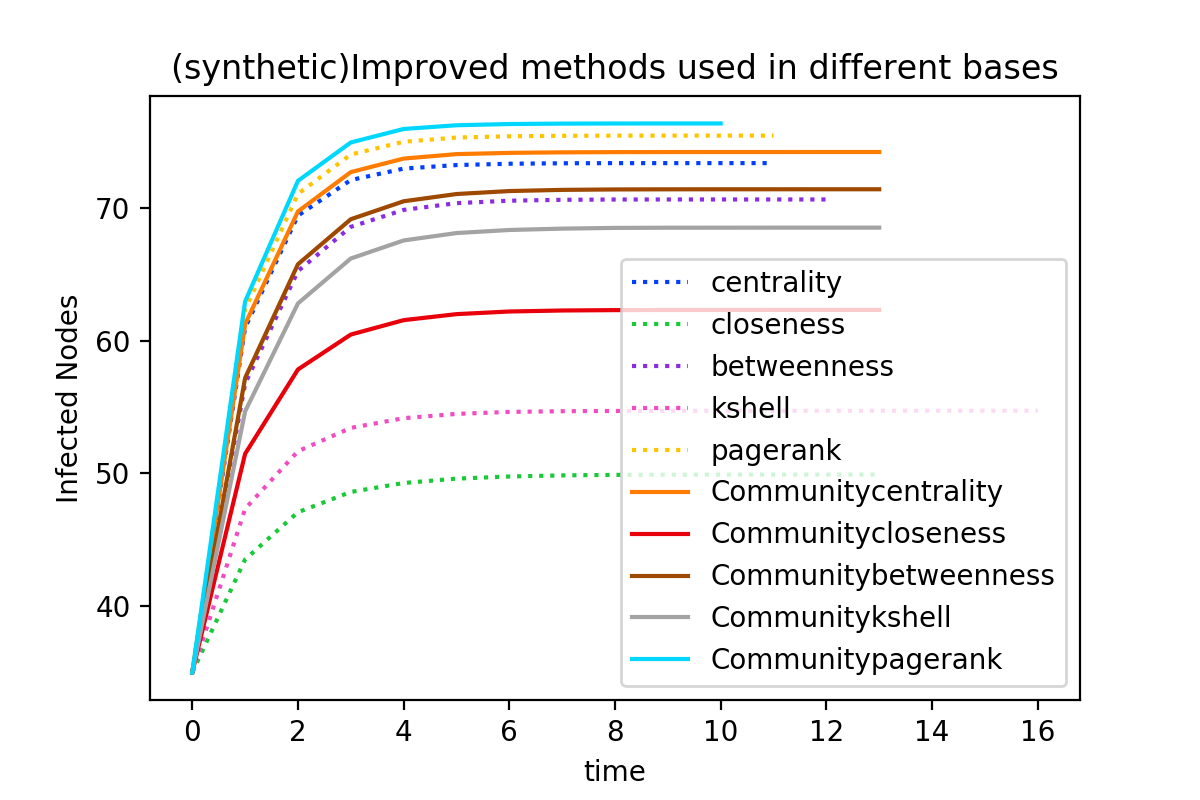}
    			\subcaption{synthetic network with 250 nodes,$\tau_1$ = 3,\\ $\tau_2$ = 1.5, $\mu$ = 0.1}
    			\label{fig:synthetic1}
    		\end{minipage}
    		\begin{minipage}[c]{0.45\textwidth}
    			\centering
    			\includegraphics[height=0.225\textheight]{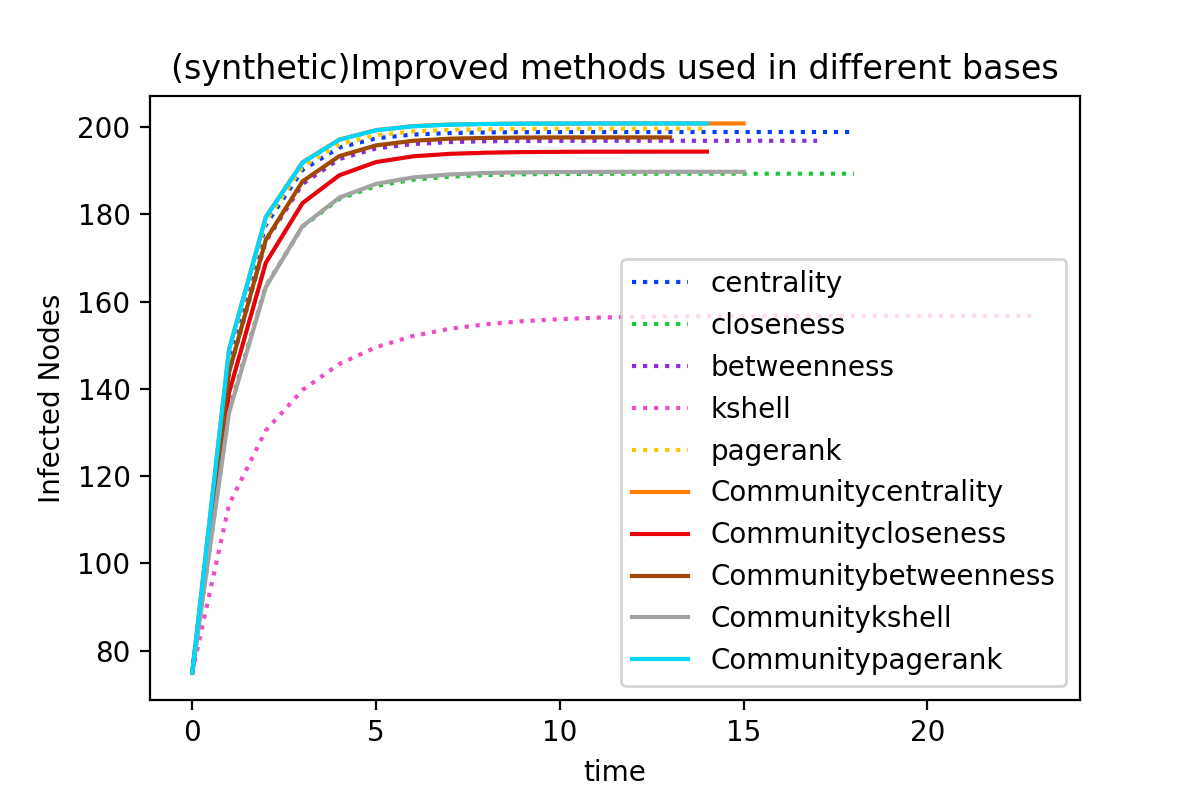}
    			\subcaption{synthetic network with 500 nodes,$\tau_1$ = 3,\\ $\tau_2$ = 1.5, $\mu$ = 0.1}
    			\label{fig:synthetic2}
    		\end{minipage}
    		\label{fig:SIR-synthetic}
\end{figure}

Two synthetic networks with different number of nodes, 250 and 500 are tested. Figure \ref{fig:SIR-synthetic} shows that with prior known community information, all five baseline ranking methods are improved respectively. CC and kshell see remarkable increase in propagation range, especially in the former network, while DC,BC and PageRank only experience a moderate growth. The results of synthetic networks are similar to that of real social network datasets.

\paragraph{Comparing different community detection methods}

In this part, different community detection methods(GN, Infomap, LPA, DEMON, K-clique and BigCLAM) are used on karate network. The Randomized Complete Block Design(RCBD) is implemented here, where node ranking methods are set as blocks, since the variability is controllable. Experiments with this factor include the Analysis of Variance(ANOVA) and pairwise comparison method Fisher Least Significant Difference(Fisher LSD). The statistical model is:$y_{ij} = \mu + \tau_i + \beta_j + \epsilon_{ij},$ where $y_{ij}$ is the sum of infected nodes, $\mu$ is the overall mean, $\tau_i$ and $\beta_j$ represent community detection methods and node ranking methods respectively.

We first test whether community information got by all community detection methods have equal improving effects to propagation result, Table \ref{tab:ANOVA} shows that different community detection methods can cause significant difference to final infected number, since p-value is 0.0069. Then, Fisher LSD test is used to find out which pairs of community detection methods cause difference. Table \ref{tab:Fisher_LSD} describes that there is significant difference between propagation result without community information and the one with results of community detection except GN. Besides, all community detection methods except GN are considered as having similar improving effect, and that of GN is proved to be similar to K-clique as well as LPA. 

The test result is that most community detection methods have less significant difference in improving maximize propagation in social networks, but the difference between with and without community information is significant at the level of $\alpha=0.05$.

\begin{table}[!ht]
\centering
\caption{The Analysis of Variance(ANOVA) Table}
\label{tab:ANOVA}
\begin{tabular}{lccccc}
\hline
Source of Variation                                                   & \multicolumn{1}{l}{Sum of Squares} & \multicolumn{1}{l}{Degrees of Freedom} & \multicolumn{1}{l}{Mean Square} & \multicolumn{1}{l}{$F_0$} & \multicolumn{1}{l}{p-value} \\ \hline
\begin{tabular}[c]{@{}l@{}}Community detection methods\end{tabular} & 1.9966                             & 6                                      & 0.3328                          & 3.9459                 & 0.0069                      \\
Node ranking methods                                                  & 8.4417                             & 4                                      & 2.1104                          &                        &                             \\
Error                                                                 & 2.0241                             & 24                                     & 0.0843                          &                        &                             \\
Total                                                                 & 12.4624                            & 34                                     &                                 &                        &                             \\ \hline
\end{tabular}
\end{table}

\begin{table*}[!ht]
\centering
\caption{Fisher LSD result}
\label{tab:Fisher_LSD}
\begin{threeparttable}
\begin{tabular}{llllllll}
\hline
Level\tnote{*}         & Infomap & DEMON   & BigCLAM & K-clique & LPA     & GN      & Base    \\ \hline
              & A       & A       & A       & A       & A       &         &         \\
              &         &         &         & B       & B       & B       &         \\
              &         &         &         &         &         & C       & C       \\
Least Sq Mean & 18.4540 & 18.4248 & 18.3738 & 18.2932 & 18.2744 & 17.9750 & 17.7610 \\ \hline
\end{tabular}
 \begin{tablenotes}
        \footnotesize
        \item[*] Levels not connected by same letter are significantly different.
      \end{tablenotes}
\end{threeparttable}
\end{table*}

\section{Discussions}

In this paper, we focused on identifying influential nodes in undirected and unweighted networks and proposed a novel method with a penalty based on uncovering both non-overlapping and overlapping community structures. The proposed method comprehensively considered the information flow transporting in the same and different communities, and the community penalized method is then put forward. In our method, we improved the information spreading by community penalty, and reduced its bias by resampling to perturb the edges of the original graph. SIR model is used here to calculate the spreading results by our proposed method, and the experimental results on real networks illustrated that it could be seen as an enhanced spreading framework for most base ranking methods. 

Our methods focus on the maximum spreading of information in complex networks, so it may better be recognized as an enhanced framework, which has been experienced to improve the epidemic spreading empirically. However, whether there is a clear principle to convince us that it can theoretically promote the base algorithms is still a problem. Besides, the experiment shows that by importing community information, most propagation results of a set of most important nodes ranking by closeness, kshell and PageRank can be improved, but that ranking by Centrality and Betweenness can not, which means that simply mixing different nodes ranking methods with community detection methods may not be appropriate enough, given their own characteristics.

Except for simulations of propagation dynamics by the SIR model, another popularly used criterion to evaluate the performance of influential nodes identification is Kendall's tau $\tau$, which measures the correlation of the ranking of the nodes. However, as it does, it only measures for the ranking instead of spreading influences, hence it is not suitable for this problem. It is still an anticipated issue to find other criteria to judge the algorithm performance.

\printbibliography

\end{document}